\documentclass[final]{IEEEcsmag}

\ifCLASSOPTIONcompsoc
  \usepackage[nocompress]{cite}
\else
  \usepackage{cite}
\fi
\ifCLASSINFOpdf
\else
\fi
\usepackage{comment}

\usepackage{physics}    
\usepackage{amssymb}    
\usepackage{amsmath}    

\usepackage{physics}

\usepackage{caption}

\usepackage{subcaption}

\usepackage{hyperref}
\usepackage{amssymb}

\usepackage{tabularx}
\usepackage[table]{xcolor}        
\usepackage{booktabs}

\usepackage{tikz}
\usetikzlibrary{shapes.geometric, arrows}
\tikzstyle{cprocess} = [rectangle, rounded corners, minimum width=4.5cm, minimum height=1cm,
                        text centered, align=center, draw=black, fill=white, font=\sffamily\footnotesize]
\tikzstyle{qprocess} = [rectangle, rounded corners, minimum width=4.5cm, minimum height=1cm,
                        text centered, align=center, draw=black, fill=gray!15, font=\sffamily\footnotesize]

\tikzset{
  arrow/.style = {thick,->,>=stealth}
}

\tikzset{
  cprocess/.style = {rectangle, rounded corners, minimum width=4.0cm, minimum height=1cm,
                     text centered, align=center, draw=black, fill=white, font=\sffamily\footnotesize},
  qprocess/.style = {rectangle, rounded corners, minimum width=4.5cm, minimum height=1cm,
                     text centered, align=center, draw=black, fill=gray!15, font=\sffamily\footnotesize},
  arrow/.style = {thick,->,>=stealth}
}

\usepackage[dvipsnames]{xcolor}

\newcommand{\mytitle}{Quantum Computing and Visualization Research Challenges and Opportunities}

\newcommand{\rev}[1]{#1}

\newcommand{\qx}{QC}

\newcommand{\insitu}{\emph{in situ}}

\newcommand{\orderof}{on the order of}
\newcommand{\hybridqcc}{HCQC}
\newcommand{\hybridqca}{HCQA}

\newcommand{\teenyvspace}{\vspace{2pt}}

\begin{document}


\title{\mytitle{}}

\author{E. Wes Bethel{\IEEEauthorrefmark{1}\IEEEauthorrefmark{2}},
Roel Van Beeumen\IEEEauthorrefmark{2}, 
Talita Perciano\IEEEauthorrefmark{2} \\
\IEEEauthorrefmark{1}San Francisco State University, 
\IEEEauthorrefmark{2}Lawrence Berkeley National Laboratory}

\begin{abstract}
 
Quantum computing (QC) has experienced rapid growth in recent years with the advent of robust programming environments, readily accessible software simulators and cloud-based QC hardware platforms, and growing interest in learning how to design useful methods that leverage this emerging technology for practical applications.
%
%
From the perspective of the field of visualization, this article examines research challenges and opportunities along the path from initial feasibility to practical use of QC platforms applied to meaningful problems.


\end{abstract}

\maketitle


%
\IEEEpeerreviewmaketitle

\section{Introduction}

Computing advances continually reshape how we think about algorithms and systems. Visualization and graphics have often been at the center of this change—most notably with the evolution of single‑purpose triangle engines into today’s general‑purpose GPUs that now power much of the AI landscape. In his 2021 Turing Award lecture, Dongarra argues that Quantum Processing Units (QPUs) will likely assume a similar role: specialized accelerators integrated into heterogeneous systems alongside CPUs and GPUs~\footnote{Online at the ACM Youtube Channel: \url{https://amturing.acm.org/vp/dongarra_3406337.cfm}, last accessed Aug. 2025.}. This prospect raises two immediate questions: how do we get there, and what are the primary challenges along the way?

This article examines those questions from a visualization perspective. We consider opportunities and challenges on the path from initial feasibility through practical use of QC platforms for useful problems. Our discussion proceeds along two complementary directions. 
First, 
we consider how a QPU might accelerate portions of a visualization workflow within a heterogeneous environment. 
\rev{This line of thinking is the primary focus of this article.}
Second, 
we consider how visualization can aid the development and use of QC itself—by improving our understanding of code and performance, algorithm structure and communication, and the high‑dimensional state spaces that arise in quantum programs.

We begin with a brief background in QC, then use a canonical visualization pipeline---isosurface extraction and rendering---as a working example for analyzing key steps and costs in hybrid classical–quantum computing settings. We discuss the practical issues of moving data from the classical to the quantum world, rethinking processing on the quantum side, and interpreting results after measurement. 
\rev{The analysis reveals several likely directions for future research needed to overcome technological and conceptual obstacles along the path towards practical use of \qx{} in visualization workloads.}


\section*{Definitions Sidebar}

\teenyvspace{}
\noindent \rev{
\emph{Quantum information} refers to information stored in the state of a quantum system, often expressed using complex probability amplitudes $\alpha \in \mathbb{C}$. 
}

\teenyvspace{}
\noindent 
\rev{
\emph{Qubit} the quantum analogue of a classical bit.
Instead of being limited to 0 or 1, its state is described by complex probability amplitudes for both $\ket{0}$ and $\ket{1}$ states, allowing it to represent a \emph{superposition} of both $\ket{0}$ and $\ket{1}$ until measured.
}

\teenyvspace{}
\noindent 
\rev{
\emph{Superposition} is a fundamental quantum property in which a quantum system can exist in multiple possible states at once. In \qx{}, this allows a computational state to represent a combination of many states simultaneously until measured.
}

\teenyvspace{}
\noindent 
\rev{
\emph{Quantum gate} is a basic operation that changes the state of one or more qubits, similar to how a logic gate acts on bits in a classical computer.
}

\teenyvspace{}
\noindent 
\rev{
\emph{Quantum scaling}  refers to how the state of $N$ qubits expands into  possible basis states. Each qubit doubles the size of the system's state space, so a 10-qubit system produces $2^{10}$ amplitudes. Because quantum operations may act on the entire state, even a single-qubit gate can change the probability amplitudes of all $2^N$ basis states when that qubit is entangled with others.
}

\teenyvspace{}
\noindent 
\rev{
\emph{Hybrid classical–quantum computing (\hybridqcc{})} describes computational systems and applications that integrate classical and quantum resources within a single heterogeneous platform, analogous to CPU–GPU systems combining host and device code. In this article, we assume such hybrid platforms and environments are ubiquitous.
}

\teenyvspace{}
\noindent 
\rev{
\emph{Hybrid classical–quantum algorithms (\hybridqca{})} integrate classical and quantum computations in iterative feedback loops—quantum processors evaluate objective functions, and classical optimizers adjust parameters—enabling the two systems to work together toward a shared solution.
}


\teenyvspace{}
\noindent 
\rev{
\emph{Error} refers to deviations in quantum computations caused by hardware imperfections and environmental noise (e.g., heat or naturally occurring radiation).
In NISQ systems, such errors—arising from decoherence (when a qubit loses its ability to maintain state), gate infidelity (when the operation on quantum state has errors), and measurement noise—accumulate over time, limiting the reliability of quantum results.
}

\teenyvspace{}
\noindent 
\rev{
\emph{Decoherence} is the process by which a qubit loses its quantum properties like superposition and entanglement because of interactions with its surrounding environment.
}


\section{Background}

\subsection{What is Quantum Computing?}

QC is a model of computation that uses the principles of quantum mechanics—such as superposition, entanglement, and interference—to process information in fundamentally different ways from classical computers~\cite{Nielsen:Quantum:2010}. 
\rev{Q}uantum computers use \emph{qubits} \rev{instead of bits. Unlike classical bits that are either 0 or 1, a qubit} can exist in a \emph{superposition} \rev{of both states, written as} $\ket{\psi} = \alpha\ket{0} + \beta\ket{1}$, where the $\ket{0}$ and $\ket{1}$ basis states are \rev{conceptually similar to the classical 0 and 1 states. 
The terms $\alpha, \beta \in \mathbb{C}$ are \emph{probability amplitudes}, satisfying $|\alpha|^2 + |\beta|^2 = 1$, whose squared magnitudes give the probabilities of measuring 0 or 1, respectively.} 

Quantum computation consists of applying \emph{unitary transformations} (quantum gates) to the quantum state $\ket{\psi}$ to produce a new quantum state $\ket{\psi'}$. To obtain the final answer from a quantum computation, the state $\ket{\psi'}$ is \emph{measured}, 
\rev{which collapses the quantum state to a basis state and produces a single bitstring. Typically, a given quantum program is run many times, each run yielding one measurement outcome. The collection of all such outcomes forms a probability distribution, revealing the most and least likely bitstrings that represent the solution to the given problem.}

\rev{Some of the key ways that QC differs from classical computing include:
(1) Representation of information — while classical bits are deterministic (always either 0 or 1), qubits can exist in a superposition of both states until measurement, yielding probabilistic outcomes; 
(2) Parallelism — superposition allows a quantum state of $N$ qubits to encode $2^N$ complex amplitudes simultaneously, enabling certain computations to explore many possibilities concurrently, even though only limited information can be extracted upon measurement;  
(3) Correlations — quantum entanglement gives rise to non-classical correlations between qubits that have no classical analog; 
(4) Algorithmic complexity — certain problems (e.g., factoring via Shor’s algorithm, unstructured search via Grover’s algorithm) exhibit  asymptotic speedups over the best known classical algorithms~\cite{Nielsen:Quantum:2010}. 
However, practical \qx{} remains limited by noise, decoherence, and the challenges of efficiently encoding and extracting data~\cite{Preskill2018quantumcomputingin}. }

\subsection{Is Quantum Computing Useful?}
\label{subsec:advantage}

\rev{This question---is \qx{} useful---has been and continues to be the subject of a significant amount of research. It turns out there is no simple answer to this question as there is a spectrum of perspectives on \emph{usefulness}.
}
%

\emph{Quantum feasibility} is a term we introduce here to \rev{describe how feasible an} operation is on QPUs. \rev{It captures the stage where executing a computational task on QC first becomes possible---and eventually practical or advantageous---}compared to classical approaches. The term \rev{serves as} an umbrella \rev{for the many challenges involved in} transition\rev{ing} from classical to quantum platforms.

\emph{Quantum utility} or \emph{quantum practicality} \rev{refers to the stage at which} a \rev{practical application, executed} on a quantum platform, \rev{requires less computing time, or less power, or yields more accurate results, compared to the best} classical \rev{device of similar size and cost}~\cite{Herrmann:2023, Hoefler:practicality:2023}. 

\emph{Quantum advantage} or \emph{Quantum supremacy}, \rev{by contrast, denotes the point where} a \qx{} outperforms classical computation on specific (not necessarily useful) tasks. For example, a quantum algorithm may \rev{achieve} quantum advantage \rev{through} significantly lower computational complexity \rev{than its} classical \rev{counterpart}~\cite{Herrmann:2023}. 


\rev{Understanding these distinctions---from feasibility to utility to advantage---is essential for assessing \qx{}'s potential in visualization and other application domains. Table~\ref{tab:feasibility_examples}} illustrates how these concepts apply to different aspects of quantum visualization research, showing the current state on Noisy Intermediate Scale Quantum (NISQ) devices, near-term prospects as fault-tolerant systems emerge, and long-term possibilities for \rev{full error-corrected fault-tolerant} quantum platforms.

\subsection{Generations of Quantum Systems}
\label{subsec:quantum_systems}

While an in-depth historical survey is beyond the scope of this article, we consider two broad generations of quantum systems to provide context for the discussion that follows.


\emph{Noisy, Intermediate-Scale Quantum (NISQ) Platforms}. Preskill (2018)~\cite{Preskill2018quantumcomputingin} introduced the term \textit{NISQ} to refer to \rev{the current generation of QCs with roughly} 50-100 qubits.
These systems are \rev{described as} \textit{noisy} \rev{because their} quantum state\rev{s are still fragile and susceptible to errors from} unintended interaction with the environment, \rev{imperfect gate operations, and decoherence. Despite their limited qubit counts and relatively short coherence times, NISQ devices provide valuable testbeds for developing algorithms, control techniques, and error-mitigation strategies that bridge the gap between proof-of-principle demonstrations and future fault-tolerant QCs}.
Present-day NISQ systems accessible via cloud-based providers typically offer \rev{\orderof{} 100} qubits and $0.01\%$ error for gates and measurement.

\emph{Scalable, Fault-Tolerant Quantum  (SFTQ) Platforms}.
This \rev{emerging} class of QCs, which \rev{are still under development but appearing} on vendor roadmaps within \rev{roughly} a 5-year horizon, \rev{aims to overcome} noise \rev{and decoherence} through \rev{quantum} error correction~\cite{Bravyi:fault-tolerant:2024}.
The \rev{core principle} is to \rev{encode a single \emph{logical qubit} using} many \emph{physical qubits} \rev{(typically 20--250), each with physical} error rates of e.g., \rev{around} $10^{-3}$, \rev{to achieve logical} error rates \rev{as low as}, e.g., $10^{-7}$.
\rev{Realizing large-scale, fault-tolerant computation is expected \rev{to require on the order of} $10^3$ to $10^9$} fault-tolerant, low-error logical qubits \rev{for demanding} algorithms like Shor's prime factorization~\cite{Nielsen:Quantum:2010}.





\section{A Visualization Example: Quantum Feasibility and Practicality}
\label{sec:vis_pipeline}

\vspace{8pt}

\subsection{Quantum Practicality in Practice}

Building on Hoefler et al., 2023~\cite{Hoefler:practicality:2023}, we examine quantum feasibility and practicality in the context of an exemplar visualization pipeline: isosurface extraction and rendering. This pipeline provides a familiar framework for \qx{} challenges, touching on issues likely to appear across a broad range of visualization workflows.


%

Hoefler et al.~\cite{Hoefler:practicality:2023} assess the prospects for practical quantum advantage and conclude that, while QCs promise speedups for certain problem classes, most current algorithms are unlikely to yield real-world benefits without major advances. To account for near-term improvements, their model assumes optimistic, better-than-current quantum performance and pessimistic classical performance. For example, Grover’s unstructured search has complexity $\mathcal{O}(\sqrt{N})$ versus $\mathcal{O}(N)$ classically, a quadratic advantage. 
Their analysis highlights a large disparity in gate-level performance—9.75 Top/s for an NVIDIA A100 versus 0.83 Kop/s for a projected large-scale error-corrected QPU. 

\rev{Their study evaluates practical quantum advantage by estimating the problem size $N$ required to achieve runtime parity between a lower-complexity quantum algorithm on a slower quantum platform and a higher-complexity classical algorithm on a faster device.
Given the 
$\approx 10^9 \dots 10^{10}$
performance gap between the
computational rates of these two platforms in their study, the problem size $N$ that results in runtime parity is quite large, resulting in a runtime on the order of weeks. }

They \rev{also} examine how this crossover point changes as the computational complexity \rev{gap widens beyond a quadratic advantage:} quantum algorithms that have a cubic or quartic advantage will shorten the time needed to achieve a crossover point for a given $N$: going past that crossover point results in practical quantum advantage.
They observe that QCs are better suited for "big compute" problems with small data rather than data-intensive applications, due to inherent input/output bandwidth limitations.
\rev{This key observation portends significant challenges for visualization, which is concerned with transformation of vast amounts of data into readily comprehensible images.}


\begin{figure}[t!]
    \centering
    \includegraphics[width=1.0\linewidth]{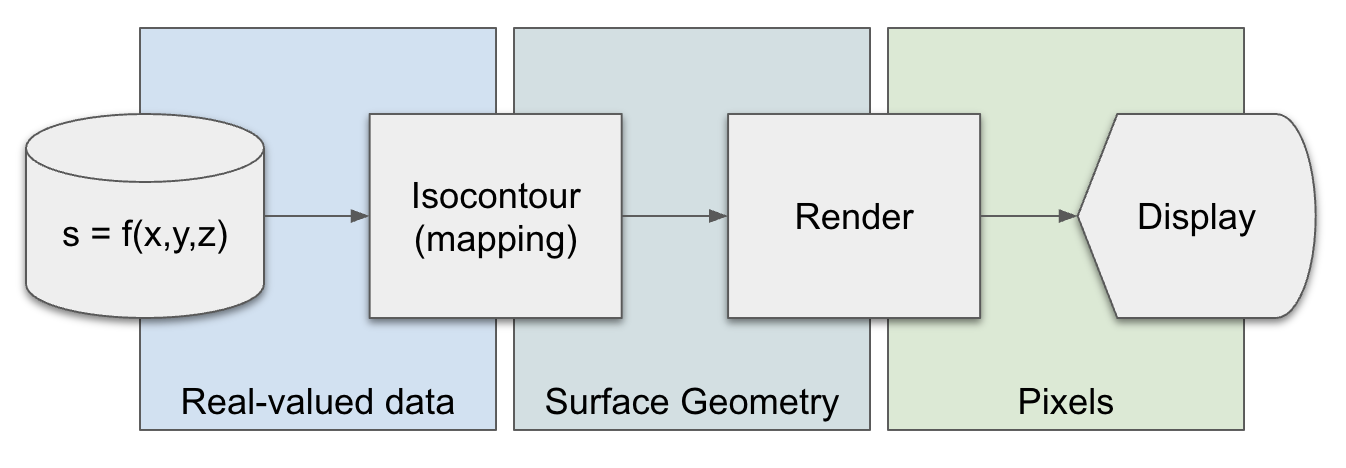}
    \caption{A canonical visualization pipeline: 3D data is input to a \emph{mapping} process, in this case isocontouring, to transform real-valued data into surface geometry, which is then input to a \emph{rendering} process that generates pixels that are then presented to a user.}
    \label{fig:iso_pipeline}
\end{figure}

\subsection{Quantum Visualization Processing}

\newcommand{\myvspace}{\vspace{0pt}}

\begin{table*}[t!]
\centering
\begin{tabularx}{\textwidth}{p{1.5cm} X X X X }
\textbf{Processing stage} & \textbf{CPU/GPU Hybrid} & \textbf{Classical performance estimate} & \textbf{CPU/Quantum hybrid} & \textbf{CPU/Quantum hybrid performance estimate} \\
\toprule
Load data & 
Load data from disk-based persistent storage into CPU/GPU RAM.  &
Disk I/O, $4N$ bytes read from disk then loaded into CPU/GPU memory. For \insitu{} settings, data may already be resident in memory. &
On the classical side: load data from disk-based persistent storage into CPU/GPU RAM; generate quantum circuit that encodes classical data into quantum state $\psi = \ket{0}^{\otimes{N}}$ &
Classical processing (same as 3rd column);
\rev{then classical creation of the quantum circuit: first, $N$ normalization operations, then generation of $\mathcal{O}(N) \dots \mathcal{O}(N^2)$ 
quantum gates for state initialization  where gate count depends on the specific encoding method (see Table~\ref{tab:encoding_costs}) } 
\\
\midrule
Isocontouring: (1) classification, (2) triangle generation & 
(1) Classification: $N$ comparison operations to compare node values to isocontour value and produce bitcodes for each mesh cell vertex; \newline
(2) Triangle generation given N bitcodes. & 
(1) $N$ memory reads, $N$ arithmetic/comparison operations, $N$ memory writes (bitstring per node); \newline
(2) \rev{$N/10$ memory reads, $(N/10)*(2.1*3)$ or $\approx N/2$ memory writes$^{a}$}.  & 
(1) Quantum classification: \rev{$\approx N$} comparison operations to classify node values; \newline
(2, Option 1.) Quantum triangle generation; or \newline
(2, Option 2.) Classical triangle generation: decode quantum state information to generate classical-format triangles.  & 
(1): in theory: \rev{ $\mathcal{O}(1){^b}$ } ;  \newline
(2, Option 1.): unknown feasibility, circuit complexity \newline
(2, Option 2.): classical readout of $N$ classification bitstrings followed by classical processing$^c$ \\
%
%
\midrule
Rendering &
Surface rendering of triangle data: local or global illumination &
\rev{$\approx N/2$ memory reads (triangle data)}; 
\rev{$M$ memory writes for an output image of $M$ pixels} &
No obvious route for use of the quantum platform, assume all rendering happens on the classical platform. &
Unknown feasibility \\
\bottomrule
\end{tabularx}
\caption{Isocontouring processing stages along with an overview of performance costs for classical and quantum implementations.
These quantum estimates are based upon a hypothetical, unspecific implementation.
$^{a}$Assumption: about $N/10$ of the cells will contain the isocontour, and each of them will result in about 2.1 triangles/cell: this estimate is highly data dependent and is based on practical experience.
\rev{$^{b}$We are assuming the possibility of applying a single operation to the entire dataset with a single gate through the property of quantum scaling.}
$^{c}$Readout of the quantum state implies the need to run the circuit many times to achieve an expected level of accuracy.
\label{table:isotable_tasks}
}
\end{table*}


We base our discussion on a canonical visualization workflow (Fig.~\ref{fig:iso_pipeline}) with mapping and rendering stages: 
\rev{a structured 3D field of size $N$ cells
is mapped (e.g., by isocontouring) into surface geometry, then rendered to an image of $M$ pixels.}
This pipeline is data-intensive and multi-stage; each stage consumes and produces different data types using different memory access patterns. Table~\ref{table:isotable_tasks} contrasts classical execution with a \rev{hypothetical hybrid CPU/QPU implementation, reflecting the practical reality that quantum methods for data-intensive problems will operate in heterogeneous settings where state preparation, execution, and measurement are combined with classical I/O and post-processing.}
%
%
Our working assumptions mirror common practice: FP32 input and INT32 pixel output, with about $N/10$ mesh cells containing the isocontour and each producing $\approx 2.1$ triangles. 
\rev{While these estimates 
are entirely data-dependent, we will adopt their use through the following discussion.}


\emph{Load data processing.}
Classically, data are read from storage (or accessed in situ). 
To prepare classical data for quantum processing, classical-side processing normalizes values and synthesizes a state-preparation circuit mapping $\ket{0}^{\otimes{N}}$ to a state encoding $N$ elements. This step is a known bottleneck: it requires \rev{$\mathcal{O}(N)$ to $\mathcal{O}(N^2)$ initialization gates with depth and gate count depending on encoding methodology} (see \S\ref{sec:data}, Table~\ref{tab:encoding_costs}). 
\emph{Isocontour generation.}
The classical Marching Cubes method~\cite{Lorenson:1987} consists of (1) a classification step where a cell's node values are compared to the isovalue to yield per-cell bitcodes, and (2) triangle generation from those bitcodes. In a \rev{hypothetical \hybridqcc{} workflow, classification could be run on the QPU, returning one bitcode per mesh cell, which is then used classically for triangle generation.}
With suitable encoding \rev{of classical data for quantum use, a quantum classification implementation of $\mathcal{O}(1)$ quantum complexity is possible using recent quantum numerical methods, thereby leveraging quantum-parallel scaling~\cite{balweski:ehands:2025};} however, current hardware requires many shots \rev{(e.g., $\approx 10^3 \dots 10^5)$} for accuracy, whereas \rev{execution on a future fault-tolerant system could reduce this to a single logical execution.} 

Performing both classification and triangle generation on the QPU faces feasibility hurdles \rev{since the data in this workflow changes type from bitstrings to triangles; this pattern does not map well to current \qx{} approaches that operate on a quantum state representing an encoding of a single type of classical data. Developing quantum-native formulations of such multi-type pipelines remains an open problem.}


\emph{Rendering.}
\rev{The performance and feasibility of classical rendering of triangles is well understood. 
From a quantum perspective, triangle rendering is a relatively unexplored topic. The quantum heterogeneous data problem identified above applies in the rendering stage as well since there is a data type change from surface geometry to pixel data. Rethinking a triangle engine for quantum hardware may be possible in the future, but feasibility is currently unknown.}


In summary, the canonical pipeline reveals the main issues any "quantum for visualization" strategy must confront: significant state-preparation costs at data ingress, mismatches at data-type transitions, and readout/accuracy constraints at egress. 
\rev{Beyond the high gate count costs associated with data encoding, factors related to mid-pipeline data type change and discovering suitable quantum implementations impact 
feasibility and practicality for data-intensive visualization on heterogeneous CPU/QPU platforms.}





\subsection{Quantum Feasibility and Practicality in Visualization}


\rev{Given the stages and costs summarized in Table~\ref{table:isotable_tasks} and the conclusions of Hoefler et al.~\cite{Hoefler:practicality:2023}, several observations follow. First, the cost of encoding classical data for quantum use---circuit synthesis and state preparation for $N$
input values---was not part of Hoefler’s model and will push the classical–quantum crossover farther out in time. This means that achieving practical quantum advantage is an even more distant target than previously thought.
Many classical data encoding methods are feasible today (see \S\ref{sec:data}) but for data‑intensive workloads like visualization they raise the bar for achieving practical utility.}

Second, when comparing quantum and classical processing costs, the feasibility–practicality balance can shift with the \rev{specific} operation. In quantum image processing, certain encodings \rev{make possible} $\mathcal{O}(1)$ unary operations across all pixels via quantum‑parallel scaling~\cite{Balewski2024}; \rev{that benefit results after an upfront encoding cost, which may be non-trivial     (\S\ref{sec:data}). }
From this perspective, filtering‑style operations may be promising \rev{candidates for leveraging quantum scaling}, while algorithms that require mid‑pipeline data‑type changes (e.g., bitstrings to triangles, triangles to images) are more difficult to map \rev{to the quantum environment} and may  yield advantage only under specific formulations and implementations.

Third, moving results back to the classical world introduces readout limits that affect utility. Finite measurement precision directly bounds the fidelity of returned values (e.g., 8‑bit readout yields only 256 distinct outcomes), and both systematic and transient errors can distort results. These effects, together with device‑level error correction and mitigation considerations, place practical constraints on accuracy until hardware and systems mature.


\section{Data Challenges}
\label{sec:data}
\vspace{2pt}

\begin{figure}[t!]
\centering
\includegraphics[width=0.99\linewidth]{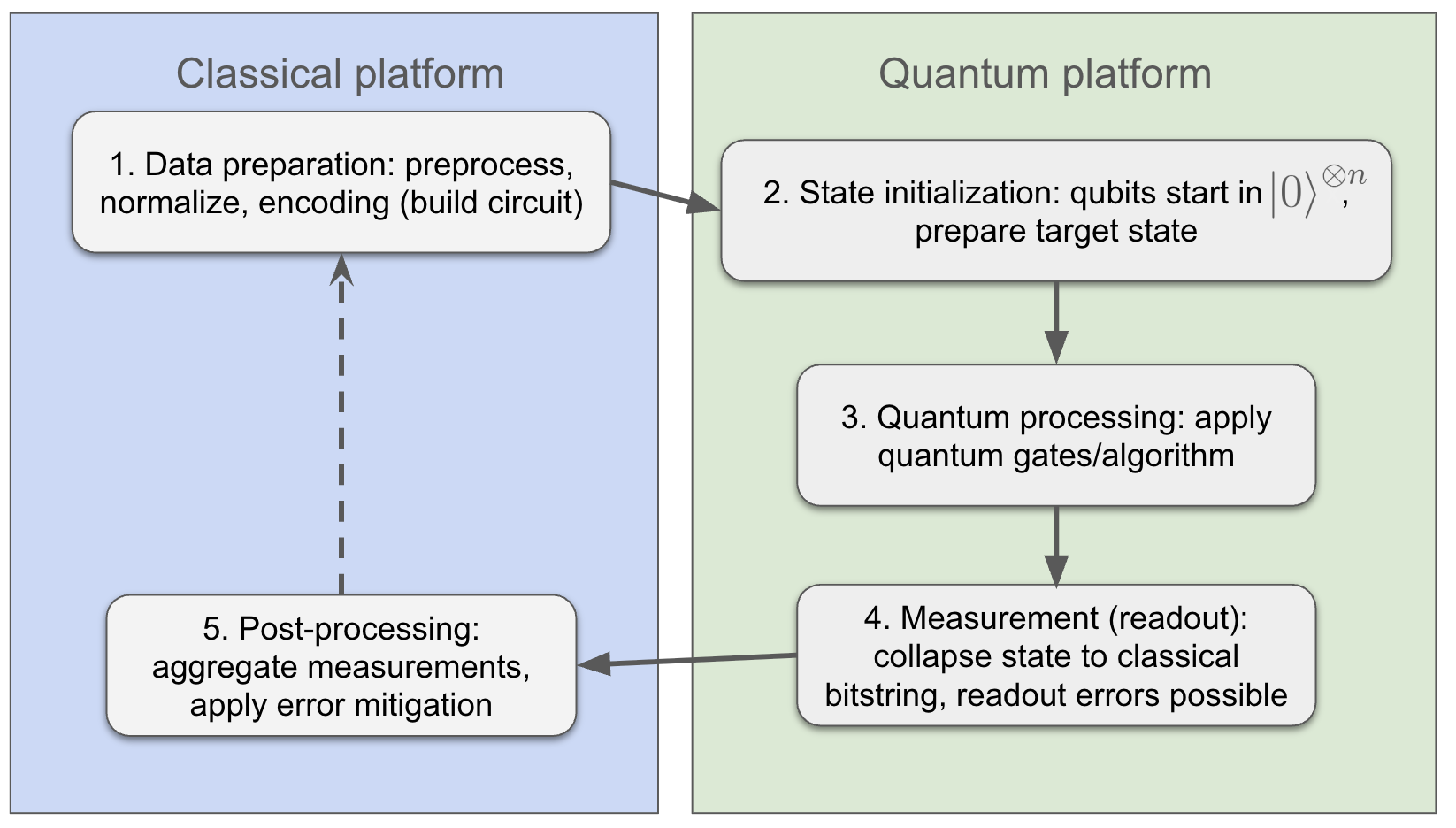}
\caption{Working with classical data on quantum platforms \rev{by definition requires} a hybrid quantum--classical workflow. The classical platform 
\rev{performs data normalization, circuit preparation, and post-processing of quantum state measurement}, 
while the quantum processor performs state initialization, quantum computation, and measurement.}
\label{fig:data_flow}
\end{figure}

Working with classical data on quantum platforms \rev{typically entails use of both classical and quantum platforms in a workflow configuration referred to as a hybrid classical-quantum computational model (\hybridqcc{}). 
This hybrid workflow, shown in Fig.~\ref{fig:data_flow}, consists of processing on both classical and quantum platforms, and data movement between them.}

\rev{
First, using one of several potential \emph{encoding} methods, the CPU transforms classical data into a form suitable for use on the QPU.
This transformation typically maps a quantum default state $\ket{0}^{\otimes N}$ to some target state $\ket{\psi}$ where the details of that mapping depend on the specific encoding method. 
Next, the QPU algorithm operates on $\ket{\psi}$, which is the quantum representation of the classical data, and produces some new result $\ket{\psi}'$, which is measured to produce a classical bitstring representation of the final answer. 
Finally, these results undergo post-processing, such as aggregation across many shots (runs of the quantum program) to produce distributions of results, as well as application of methods to mitigate errors. 
 As shown in Fig.~\ref{fig:data_flow}, this workflow introduces costs in data movement, circuit compilation, and iterative optimization—factors as important as gate counts when assessing feasibility.
}


\begin{table*}[ht!]
\centering
\caption{Comparison of quantum data encoding schemes for $N = 2^n$ data values. Here $n$ denotes the number of address qubits and $n_d$ represents the number of data qubits.}
\label{tab:encoding_costs}
\begin{tabular}{l|c|c|c|c|c}
\toprule
\textbf{Encoding} & \textbf{Qubits} & \textbf{Circuit} & \textbf{Gate} & \textbf{Classical} & \textbf{Measurement} \\
\textbf{Method} & \textbf{Required} & \textbf{Depth} & \textbf{Count} & \textbf{Preprocessing} & \textbf{Shots} \\
\midrule
\textbf{Basis}~\cite{Schuld:MLQC:2021} & $N$ & $\mathcal{O}(N)$ & $\mathcal{O}(N^2)$ & Minimal & $\mathcal{O}(N)$ \\
\textbf{Amplitude}~\cite{Schuld:MLQC:2021} & $\log_2(N)$ & $\mathcal{O}(N)$ & $\mathcal{O}(N^2)$ & Normalization & $\mathcal{O}(2^n)$ \\
\textbf{Phase}~\cite{Schuld:MLQC:2021} & $\log_2(N)$ & $\mathcal{O}(\log N)$ & $\mathcal{O}(N)$ & Problem-dependent & Algorithm-specific \\
\textbf{FRQI}~\cite{Le2011} & $n + 1$ & $\mathcal{O}(N)$ & $\mathcal{O}(N^2)$ & Minimal & $\mathcal{O}(10^4)$ per pixel \\
\textbf{QPIXL}~\cite{Amankwah:NatureScientificReports:2022} & $n + 1$ & $\mathcal{O}(\log N)$ & $\mathcal{O}(N)$ & $\mathcal{O}(N \log N)$ & $\mathcal{O}(10^4)$ per pixel \\
\textbf{QCrank}~\cite{Balewski2024} & $n + n_d$ & $\mathcal{O}(\log N)$ & $\mathcal{O}(N)$ & $\mathcal{O}(N \log N)$ & $\mathcal{O}(10^6)$ total \\
\textbf{QBArt}~\cite{Balewski2024} & $n + n_d$ & $\mathcal{O}(\log N)$ & $\mathcal{O}(N)$ & $\mathcal{O}(N \log N)$ & $\mathcal{O}(10^2)$ per value \\
\bottomrule
\end{tabular}
\end{table*}


\emph{Encoding \rev{methods}.} 
Classical values must be normalized and embedded in a quantum state before \rev{quantum} computation. Standard methods include basis encoding (direct bitstring mapping to $\ket{\psi}$~\cite{Nielsen:Quantum:2010}; simple, no compression), amplitude encoding (packs $2^N$ values into $N$ qubits~\cite{Nielsen:Quantum:2010, Schuld:MLQC:2021}; powerful but deep circuits), and phase encoding (encodes data in relative phases; sometimes shallower circuits but less intuitive~\cite{Schuld:MLQC:2021}). For images and \rev{structured} fields, the QPIXL encoding \rev{unifies pixel representations by providing a general framework that encompasses multiple quantum image representation methods~\cite{Amankwah:NatureScientificReports:2022}}, reducing circuit complexity from $\mathcal{O}(N^2)$ to $\mathcal{O}(N)$ for $N$ pixels, eliminating ancilla qubits and enabling Walsh–Hadamard compression.
\rev{The state preparation circuit has linear complexity $\mathcal{O}(N)$ and consists of $N$ $R_y$ gates and $N$ CNOT gates.}
More recent quantum-parallel encodings include QCrank (real-valued rotations via uniformly controlled gates) and QBArt (binary encoding with fewer measurements and established arithmetic)~\cite{Balewski2024}. Table~\ref{tab:encoding_costs} compares these methods by qubits, depth, gate count, preprocessing, and measurement shots.



\emph{Practical dataset and device limits.} 
\rev{From a practicality perspective, there is a vast gap between typical classical dataset sizes and those that can be handled on current quantum platforms. Consider a \emph{smaller-sized} $512^3$ volume, which consists of $\approx1.34 \times 10^8$ voxels, requiring 128 MiB of classical memory storage assuming 8-bit voxels. 
An efficient quantum encoding method can represent such a volume using $\log_2(N)$ address qubits plus data qubits -- in this case, $\log_2(2^{27}) = 27$ address qubits and 1 data qubit for a QPIXL-FRQI encoding ($N=2^{27}$ pixels total, representing the $512^3 = 2^9\times 2^9\times 2^9$ voxels in a flattened representation).
The resulting state preparation circuit would require $2^{27}$ (approximately 134 million) $R_y$ gates and an equal number of CNOT gates.
However, 
present-day NISQ devices impose other limits: they consist of $\approx 150$ qubits, and have gate/readout error rates of $10^{-2} \dots 10^{-3}$. As a result, usable circuit depths must be $<100$ before noise dominates. }

\rev{To date, demonstrations of quantum methods for data-intensive workloads like quantum image processing are limited to data sizes of only tens to hundreds of pixels 
underscoring that encoding cost, which can produce deep and wide circuits, along with device fidelity, has a greater impact on feasibility than qubit count alone. 
In order to accommodate classical dataset sizes larger than "toy problems", the error rates on quantum platforms must be significantly lower than present-day NISQ systems.
Such lower error rates are anticipated on future Scalable, Fault-Tolerant Quantum (SFTQ) Platforms. }




\emph{Measurement and readout.}
\rev{
Results from the quantum computation are obtained through 
measurement governed by the Born rule~\cite{Nielsen:Quantum:2010}, which links the quantum state's amplitudes to classical outcome probabilities: if a system in state $\ket{\psi}$ is measured in basis $\{ \ket{i} \}$, the probability of observing outcome $i$ is $|\langle{i}|\psi\rangle|^2$.
Each measurement collapses the quantum state into a classical bitstring sampled from this distribution. Because readout is inherently noisy, circuits must be executed repeatedly (many "shots"), and the resulting counts are aggregated using statistical estimation and error-mitigation methods such as detector tomography, maximum-likelihood estimation, or readout calibration~\cite{BlumeKohout2020volumetricframework}.
The required shot count can be substantial: with $N_a$ address qubits, full state recovery scales with $2^{N_a}$ measurement outcomes, and achieving $\approx 1\%$ statistical precision may require $\approx 10^6$ repetitions (shots). 
At the device level, readout fidelity is limited by state-preparation-and-measurement (SPAM) errors, finite signal-to-noise ratios in the amplification chain, and crosstalk among simultaneously measured qubits. Many platforms (e.g., superconducting qubits and trapped ions) have limited parallel readout bandwidths, requiring sequential or multiplexed measurement that constrains throughput~\cite{chen2023transmon}. Readout time, typically microseconds to milliseconds per shot, can therefore dominate total runtime in data-intensive workloads. 
}




\emph{End-to-end implications.} The classical–quantum interface introduces overheads beyond the quantum processing unit (QPU) itself: host–device communication latency, circuit synthesis and compilation time, readout and postprocessing costs, and the computational expense of error mitigation. These factors are strongly application-dependent and become favorable only when the quantum stage performs substantial computation per encoded datum and produces low-dimensional classical outputs. In this regime, encoding and measurement costs are effectively amortized, potentially making hybrid classical–quantum pipelines competitive for specific data-intensive and visualization-oriented workloads.
Such trade-offs in the classical–quantum boundary have been observed in hybrid workflows for scientific computing and machine learning~\cite{Cranganore2024FGCS}.




\section{QC and Visualization Algorithms}
\label{sec:visalgs}
\vspace{2pt}

\rev{Classical visualization algorithms typically follow a pipeline of filtering, mapping, and rendering (Fig.~\ref{fig:iso_pipeline}). These stages are computationally modest; most scale linearly or at worst quadratically with input size even on large datasets. Filtering selects and preprocesses data; mapping transforms fields into geometric representations such as isosurfaces or volumes; and rendering produces pixels for display. While advanced techniques (e.g., global illumination or physically based rendering) increase computational cost, the dominant pattern couples high data volume with relatively simple, data-parallel operations.}

\rev{
Quantum algorithms, by contrast, manipulate the system’s quantum state 
$\ket{\psi}$ through superposition and interference, amplifying desired outcomes while suppressing others. This paradigm favors high-complexity, low-data problems, exemplified by Shor’s factorization algorithm and Grover’s search \cite{Nielsen:Quantum:2010}. }

\rev{This contrast highlights a fundamental mismatch with visualization workloads, which are inherently data-intensive. A typical workflow ingests 
$\mathcal{O}(N^3)$ field values and produces $\mathcal{O}(N^2)$ pixels through straightforward, highly parallel steps.
Such patterns fall outside the small-data, high-complexity regime where quantum algorithms are most effective, making direct quantum implementations of classical visualization pipelines unlikely to provide practical advantage.}




\rev{Recent advances in quantum numerical methods suggest new avenues for data-intensive visualization. Protocols for polynomial evaluation, such as EHands~\cite{balweski:ehands:2025}, demonstrate $\mathcal{O}(1)$ quantum complexity for computing polynomial functions on real-valued encodings. If portions of visualization pipelines can be reformulated as polynomial computations or related linear and spectral transforms, quantum resources may provide utility particularly in regimes where computational intensity dominates I/O and measurement overhead.}



\rev{
Realizing such opportunities requires rethinking visualization algorithm and workflow architecture. Rather than a direct translation from classical into quantum form,
new designs should focus on identifying quantum-amenable kernels while minimizing both encoded input and measured output. Promising directions may include quantum-enhanced feature detection on high-dimensional fields that yield compact descriptors for subsequent classical rendering, and optimization or pattern-recognition tasks that leverage quantum parallelism. These strategies depart from direct geometric processing and align more naturally with the intrinsic strengths of quantum computation. In practice, achieving this alignment will depend on continued progress in data encoding, error mitigation, and scalable hybrid execution.}



\rev{
Hardware constraints remain a significant limiting factor. 
Current NISQ devices impose limits on both qubit counts and circuit depth, restricting algorithms and implementations to small problems due to error rates and limited fidelity. 
Consequently, near-term progress will most likely occur in 
hybrid pipelines where QPUs address computational bottlenecks while CPUs and GPUs manage data movement, visualization, and rendering. As SFTQ architectures mature and support larger logical-qubit counts, more ambitious applications will become feasible, potentially enabling tighter integration of quantum computations within interactive visualization workflows. Realizing this vision will require algorithms that map visualization subproblems onto quantum models while preserving the low-latency interactivity essential for scientific exploration.
}

\begin{figure*}[t!]
     \begin{subfigure}[b]{0.46\textwidth}
         \centering
         {\includegraphics[width=\textwidth]{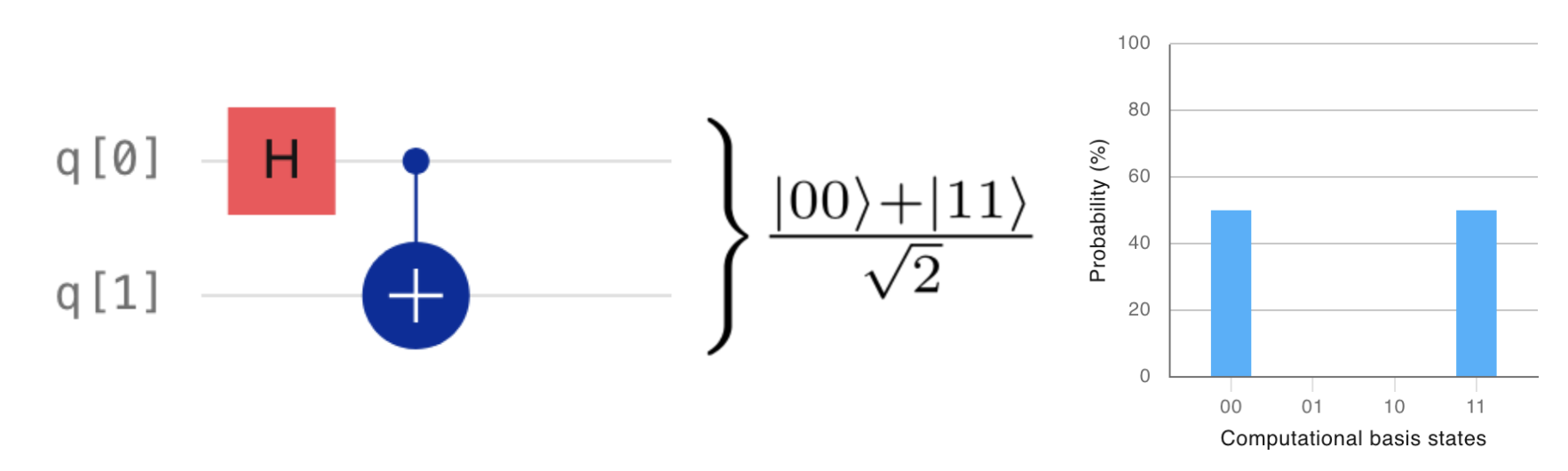}}
         \caption{\rev{A} Bell State~\cite{Nielsen:Quantum:2010} result\rev{ing} from the combination of superposition and entanglement of two qubits.}
         \label{fig:bell_state}
     \end{subfigure}
     \hfill
     \begin{subfigure}[b]{0.46\textwidth}
         \centering
         \raisebox{0.1in}
         {\includegraphics[width=\textwidth]{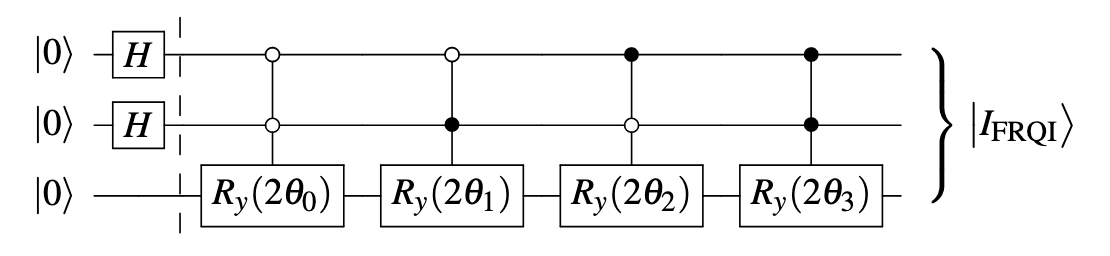}}
         \caption{A quantum circuit implementing FRQI encoding for a $2 \times 2$ pixel image with the pixel values given by $\theta_i$. Image source~\cite{Amankwah:NatureScientificReports:2022}.}
        \label{fig:ifrqi}
     \end{subfigure}
     \caption{In \qx{}, entanglement occurs when the state of two qubits is linked: change in the state of one is "seen" by the other. Both cases show circuits that leverage entanglement and superposition to achieve actions in the circuit that are not possible in any classical sense where a change in one qubit's state affects others.}
          \label{fig:entanglement}
\end{figure*}

\section{Visualization for Quantum Computing}
\label{sec:vis4quantum}
\vspace{4pt}

\rev{
Classical-side visualization for quantum computing generally falls into three categories: quantum state, circuit topology, and performance.
For single-qubit systems, the Bloch sphere~\cite{Nielsen:Quantum:2010}
provides an intuitive representation of probability amplitudes and phase. The Q-sphere method\footnote{IBM Quantum Composer User Guide, online at \url{https://quantum-computing.ibm.com/composer/docs/iqx/} last accessed Nov. 2025.} extends this concept to small multi-qubit systems by displaying amplitudes and relative phases across basis states, though it remains practical only for very small numbers of qubits ($N \leq 5$). 
Circuit-topology diagrams visualize gate sequences and state flow, revealing where entanglement is introduced (Fig.~\ref{fig:entanglement}) but not how it evolves.
Performance visualizations, such as volumetric benchmark plots~\cite{BlumeKohout2020volumetricframework}, map circuit width (qubits) and depth (gates) to achievable reliability, enabling comparative assessment across devices and calibration regimes.
From an algorithmic performance standpoint, the VIOLET system~\cite{Ruan:Violet:2024} illustrates how tunable algorithmic parameters influence the behavior and convergence of quantum neural networks through interactive visual presentation.}

\rev{
\emph{Data size challenge.} 
Visualization methods that depict state explicitly won't scale to current QPUs, which have $\approx 150$ qubits and the corresponding state space of $2^{150}$ complex probability amplitudes.
By comparison, even the largest classical platforms provide only a few  PiB of memory -- $2^{52}$ values -- leaving a substantial gap that will widen as vendor roadmaps push toward systems with $10^3$ or more logical, fault-tolerant qubits. 
Bridging this gap will require visualization approaches that convey structure, behavior, and correlations without resorting to exhaustive state enumeration.}

\rev{
\emph{Data complexity (entanglement) challenge.}
Quantum programs rely on superposition, interference, and entanglement, producing correlations and interactions that are non-local, dynamic, and difficult to visualize. Circuit diagrams indicate where entanglement is introduced but reveal little about its strength, structure, or temporal evolution. Preskill’s concept of the "entanglement frontier" delineates the regime in which correlations become so numerous and intricate that classical simulation is no longer feasible~\cite{Preskill2018quantumcomputingin}. This frontier underscores the need for visualization methods that convey relational structure and dynamical behavior without requiring full state reconstruction. Addressing this need motivates a line of research we term \emph{quantum visualization}, distinct from classical approaches that scale only to the smallest systems.
}




\section{Sidebar: Summary of Research Challenges}
\label{sec:challenges}
\vspace{2pt}

\emph{Classical–quantum data interface.}
There is a significant cost associated with quantum processing of classical data in terms of preprocessing, building the circuit to create the quantum state that encodes the classical data, and measurement of results. This cost can be an impediment to realizing quantum utility and practical quantum advantage.
More work is needed to find ways to reduce the cost and complexity of encoding and measurement while preserving the fidelity of scientific data.

\emph{\rev{Native} quantum visualization methods.}
Classical approaches for visualization algorithms are not likely to map well to quantum platforms because they do not take advantage of quantum characteristics like quantum scaling, superposition, and entanglement, nor do they align with the "small data, large compute" paradigm that does work well on quantum platforms.
Significant research is needed to find approaches that leverage the QPU to accelerate visualization workloads in whole or in part.


\emph{Limits of QC hardware.}
Present-day quantum platforms remain limited by small numbers of noisy qubits, restricted circuit depth, and high operational cost. These constraints mean that many visualization algorithms are not yet feasible to run on quantum hardware. While vendor roadmaps point toward fault-tolerant systems with thousands of logical qubits, those systems are still years away. Understanding what is possible in the near term, and preparing methods that will be ready when larger-scale platforms arrive, is a central challenge for the field.

\emph{Visualizing quantum state.}
Central challenges include visual representation of quantum phenomena like entanglement and superposition (Fig.~\ref{fig:entanglement}) in terms of how they impact quantum state, algorithm performance, and results. The characteristics of quantum scaling also present large-data challenges that have the potential to far exceed the capabilities of today's largest classical platforms. 
Developing new methods that help us see and reason about complex patterns of entanglement, interference, and state evolution will be essential if visualization is to play a role in advancing our understanding of large-scale quantum systems.

\emph{Feasibility and utility for visualization workloads.}
Even though significant barriers exist today—encoding costs, limited qubit counts, and error-prone systems—it is important to continue exploring how visualization and quantum computing can come together. Progress in hardware, algorithms, and software environments is steady, and each step opens new possibilities. By working now to understand feasibility and utility, the visualization community will be positioned to take advantage of the opportunities that fault-tolerant quantum systems are expected to provide in the years ahead.

\begin{table*}[htbp]
\centering
\caption{\rev{Current and near-term prospects for quantum computing in visualization, categorized by feasibility (what is possible today), utility (what may provide practical benefits), and advantage (what may eventually outperform classical approaches). Examples are based on current research and realistic projections.}}
\label{tab:feasibility_examples}
\small
\begin{tabular}{p{2.6cm}|p{3.8cm}|p{3.8cm}|p{4cm}}
\toprule
\textbf{Research Area} & \textbf{Current Feasibility (NISQ Era)} & \textbf{Near-Term Utility (Early SFTQ)} & \textbf{Long-Term Advantage (Mature SFTQ)} \\
\midrule
\textbf{Data Encoding} & 
\begin{list}{$\bullet$}{\setlength{\itemsep}{0pt}\setlength{\parsep}{0pt}\setlength{\topsep}{0pt}\setlength{\partopsep}{0pt}\setlength{\leftmargin}{1em}\setlength{\labelwidth}{0.5em}\setlength{\labelsep}{0.3em}}
\item QCrank/QBArt encoding of small datasets (hundreds of pixels) on real hardware~\cite{Balewski2024}
\item Proof-of-concept quantum state preparation for scientific image data with compression capabilities
\end{list}
 &
\begin{list}{$\bullet$}{\setlength{\itemsep}{0pt}\setlength{\parsep}{0pt}\setlength{\topsep}{0pt}\setlength{\leftmargin}{1em}\setlength{\labelwidth}{0.5em}\setlength{\labelsep}{0.3em}}
\item Efficient quantum representations for specific data types with compression
\item Reduced classical preprocessing for certain encoding schemes
\end{list}
 &
\begin{list}{$\bullet$}{\setlength{\itemsep}{0pt}\setlength{\parsep}{0pt}\setlength{\topsep}{0pt}\setlength{\leftmargin}{1em}\setlength{\labelwidth}{0.5em}\setlength{\labelsep}{0.3em}}
\item Potential advantages for high-dimensional data when combined with quantum algorithms
\item Theoretical storage benefits require practical quantum memory
\end{list} 
 \\
\midrule
\textbf{Quantum for Visualization} & 
\begin{list}{$\bullet$}{\setlength{\itemsep}{0pt}\setlength{\parsep}{0pt}\setlength{\topsep}{0pt}\setlength{\leftmargin}{1em}\setlength{\labelwidth}{0.5em}\setlength{\labelsep}{0.3em}}
\item Proof-of-concept quantum operations on encoded data
\item Small-scale feature detection demonstrations
\item Hybrid parameter optimization for simples cases
\end{list}
 &
\begin{list}{$\bullet$}{\setlength{\itemsep}{0pt}\setlength{\parsep}{0pt}\setlength{\topsep}{0pt}\setlength{\leftmargin}{1em}\setlength{\labelwidth}{0.5em}\setlength{\labelsep}{0.3em}}
\item Quantum-enhanced optimization for adaptive visualization parameters
\item Specialized quantum kernels for high-complexity, low-data operations
\end{list}
 &
\begin{list}{$\bullet$}{\setlength{\itemsep}{0pt}\setlength{\parsep}{0pt}\setlength{\topsep}{0pt}\setlength{\leftmargin}{1em}\setlength{\labelwidth}{0.5em}\setlength{\labelsep}{0.3em}}
\item Polynomial speedups for specific visualization subproblems (e.g., feature extraction)
\item Acceleration possible when computation dominates I/O costs
\end{list}
 \\
\midrule
\textbf{Visualization for Quantum} & 
\begin{list}{$\bullet$}{\setlength{\itemsep}{0pt}\setlength{\parsep}{0pt}\setlength{\topsep}{0pt}\setlength{\leftmargin}{1em}\setlength{\labelwidth}{0.5em}\setlength{\labelsep}{0.3em}}
\item Circuit topology visualization (existing tools)
\item Small system state visualization (Bloch sphere, Q-sphere for $N \leq 5$)
\item Performance benchmarking plots
\end{list}
 &
\begin{list}{$\bullet$}{\setlength{\itemsep}{0pt}\setlength{\parsep}{0pt}\setlength{\topsep}{0pt}\setlength{\leftmargin}{1em}\setlength{\labelwidth}{0.5em}\setlength{\labelsep}{0.3em}}
\item Improved visualization tools for debugging quantum programs
\item Interactive explorations of moderate-size quantum states
\item Real-time algorithm performance visualization
\end{list}
 & 
\begin{list}{$\bullet$}{\setlength{\itemsep}{0pt}\setlength{\parsep}{0pt}\setlength{\topsep}{0pt}\setlength{\leftmargin}{1em}\setlength{\labelwidth}{0.5em}\setlength{\labelsep}{0.3em}}
\item Novel visualization methods for entanglement structure
\item Techniques for conveying correlations without full state enumeration
\item Quantum visualization as a research field
\end{list}
 \\
\bottomrule
\end{tabular}
\end{table*}

\section{Conclusions}
\label{sec:conclusions}
\vspace{2pt}

The field of \qx{} is undergoing a rapid technological evolution that spans improvements in quantum platforms along with the software ecosystem for using them.
As part of this trajectory, we can reasonably anticipate that the current gaps between classical and quantum systems in computational rates and computation quality will narrow over time. Like Dennard scaling, technological innovations will remedy some of this performance differential over time.

Meanwhile, \qx{} offers advantages unattainable classically, enabled by phenomena such as quantum scaling, superposition, and entanglement. Terms like quantum supremacy and practical advantage mark milestones of successful application, but quantum feasibility provides a stepping stone for fields like visualization to prepare for quantum methods and platforms. Key issues surface when analyzing a canonical visualization pipeline, which is highly data-centric and requires transitions between diverse data types and representations.


\section{Acknowledgment}
\vspace{2pt}

\rev{This work was supported by the U.S. Department of Energy (DOE), Office of Science, Office of Advanced Scientific Computing Research (ASCR) under Award Number DE-SC0025382, program manager Dr. Marco Fornari, and under Contract No. DE-AC02-05CH11231.
The authors also wish to acknowledge numerous discussions with Dr. Janine Bennett (Sandia National Laboratories) that have helped to shape the scope and content of this article.
}






















\ifCLASSOPTIONcaptionsoff
  \newpage
\fi

\bibliographystyle{IEEEtran}
\bibliography{main}

\begin{IEEEbiography}{E. Wes Bethel}{\,}is a Professor of Computer Science at \emph{San Francisco State University} and a Research Affiliate at \emph{Lawrence Berkeley National Laboratory}. Contact him at ewbethel@sfsu.edu. 
\end{IEEEbiography}




\begin{IEEEbiography}{Roel Van Beeumen}{\,}is a Staff Scientist at \emph{Lawrence Berkeley National Laboratory}. Contact him at rvanbeeumen@lbl.gov. 
\end{IEEEbiography}



\begin{IEEEbiography}{Talita Perciano}{\,}is a Research Scientist at \emph{Lawrence Berkeley National Laboratory}. Contact her at tperciano@lbl.gov. 
\end{IEEEbiography}



\end{document}